# Protein Energy Landscapes Determined by 5-Dimensional Crystallography

Marius Schmidt[*,†], Vukica Srajer[‡], Robert Henning[‡], Hyotcherl Ihee[∥], Namrta Purwar[†], Jason Tenboer[†], Shailesh Tripathi[†]

[†]Physics Department, University of Wisconsin-Milwaukee, Milwaukee, WI, USA,

[‡]Center for Advanced Radiation Sources, The University of Chicago, Chicago, IL, USA

[∥]Nanomaterials and Chemical Reactions and Institute for Basic Science, Daejeon, 305-701, Republic of Korea and Department of Chemistry, KAIST, Daejeon, 305-701, Republic of Korea

ABSTRACT: Free energy landscapes decisively determine the progress of enzymatically catalyzed reactions[1]. Time-resolved macromolecular crystallography unifies transient-state kinetics with structure determination[2-4] because both can be determined from the same set of X-ray data. We demonstrate here how barriers of activation can be determined solely from five-dimensional crystallography[5]. Directly linking molecular structures with barriers of activation between them allows for gaining insight into the structural nature of the barrier. We analyze comprehensive time series of crystallographic data at 14 different temperature settings and determine entropy and enthalpy contributions to the barriers of activation. 100 years after the discovery of X-ray scattering, we advance X-ray structure determination to a new frontier, the determination of energy landscapes.

## Introduction:

Since the 1890s the Van't Hoff-Arrhenius equation, $\nu \exp^{-\beta E_a}$, has been used to describe the temperature dependence of chemical reaction rates. $E_a$ is the energy of activation and the factor $\beta = 1/(k_B T)$ containing the Boltzmann factor $k_B$ accounts for the inverse temperature behavior. The pre-factor ν accounts for the dynamic behavior of the members of the ensemble. Eyring[6] tied this equation to a transition state at the top of the barrier of activation

$$k = \frac{RT}{N_A h} e^{\frac{\Delta S^{\#}}{R}} e^{-\frac{\Delta H^{\#}}{RT}} \qquad (1)$$

where R is the gas constant, $N_A$ Avogadro's number, h Planck's constant and $\Delta S^{\#}$ and $\Delta H^{\#}$ the entropy and enthalpy differences from an initial state to the transition state, respectively. A reaction can be followed with time-resolved methods, from which conclusions on the underlying mechanism are drawn by kinetic modeling. In all early approaches[7,8], the structures of the reaction intermediates were inferred from static crystallography. Time-resolved crystallography[9] (TRX) finally unified kinetics with structure determination[10,11]. Once the structures of intermediates are known, kinetic mechanisms can be tested by post-refinement against the TRX data[4]. Microscopic rate coefficients $k_i$ between the intermediates plus the extent of reaction initiation specify a mechanism. The $k_i$ are temperature dependent which can be described by the transition state equation (TSE, eq. 1). If the temperature is varied, the previously 4-dimensional crystallographic data become 5-dimensional (5D)[5]. The photocylce of photoactive yellow protein (PYP) features a number of intermediate states, structures of which were determined earlier with picosecond TRX at only one temperature[12,13]. The nomenclature for intermediates used in previous studies[12,14-17] is followed here. Absorption of a blue photon at 485 nm provides 245 kJ/mol of energy to excite the central p-coumaric acid (pCA) chromophore (Fig. 1). Part of the energy is rapidly dissipated[18]. The remaining energy is stored in an energy rich atomic configuration[19] labeled $I_T$. The chromophore is not yet fully isomerized from *trans* to *cis*[12,13,20]. The $I_T$ state is followed by two states, $I_{CT}$ and $pR_1$. $I_{CT}$ and $pR_1$ are fully *cis*, and branch away from $I_T$ in a volume conserving bicycle-pedal and hula-twist reaction, respectively[12]. The dominant species is $I_{CT}$. In $I_{CT}$ the carbonyl-oxygen is flipped to the other side but the chromophore head is still fixed by two hydrogen bonds to amino acids Tyr42 and Glu46. In $pR_1$ the chromphore head hydroxyl lost one hydrogen bond. The entire chromphore has rotated about the chromophore axis. $I_{CT}$ relaxes to $pR_2$. This relaxation causes the Cys69 sulfur, to which the chromophore is bound, to move significantly. The strongest difference electron density features are found near this sulfur. States $pR_1$ and $pR_2$ are occupied many orders of magnitude in time. Finally, they relax to the pB state[15,21]. The pB state most likely resembles the signaling state of PYP. The chromphore head forms new hydrogen bonds with the displaced Asp52 and with an additional water that appears near the entrance to the chromphore pocket[22]. Finally, pB relaxes to the dark state (pG). Although the structures of the intermediates are now known,

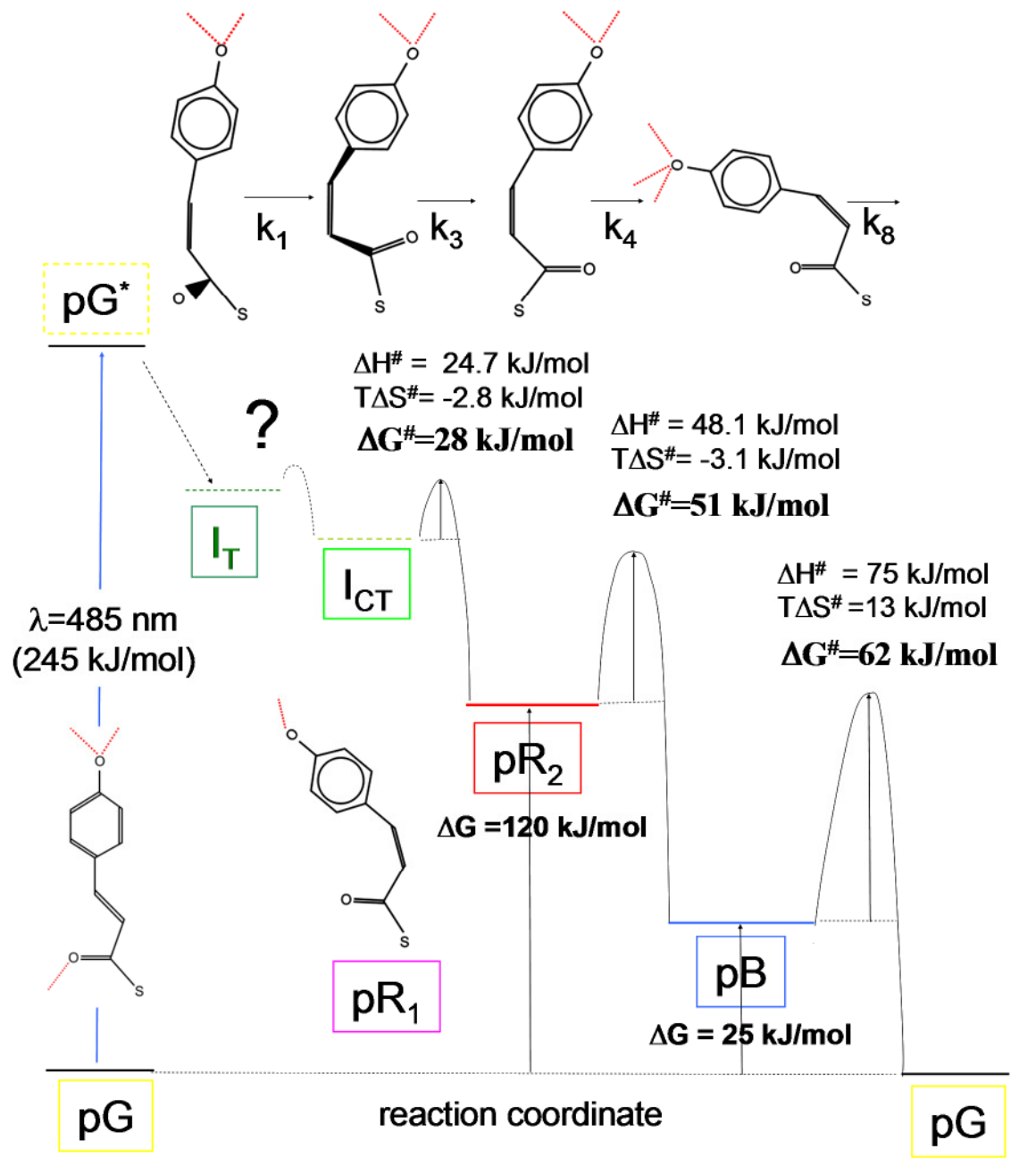

**Figure 1**. Free energy landscape of the main reaction pathway in PYP. The reaction coordinate has periodic boundaries (pG). Atomic structures of pG, $I_T$ and $I_{CT}$, $pR_1$, $pR_2$ and pB are known; early transitions from $pG^*$ require higher time-resolution; energy levels of $I_T$ and $I_{CT}$ are unknown (dashed lines); free energy levels of pR and pB are known in solution. The free energy and entropic and enthalpic contributions to the barriers as extracted from the 5D crystallographic data are shown at 300 K

information on barriers of activation in the photocycle is sparse. Here we demonstrate how these thermodynamic parameters can be extracted solely from 5D crystallography.

## Material and Methods:

Detailed information on methods is provided in the Supporting Information (SI). TRX experiments were conducted on crystals of PYP at BioCARS (Advanced Photon Source). Nanosecond laser pulses were used to initiate the photocycle and the reaction was followed from nanoseconds to seconds. The temperature was varied between -40 °C and +70 °C. The photocycle was probed at 14 different temperature settings using 21-31 time points, each. The TRX data were analyzed by Singular Value Decomposition (SVD) using the program SVD4TX[11]. Time-dependent concentrations were calculated using the program GetMech[21], which was used to optimize the microscopic rate coefficients, $k_i$, by fitting calculated time-dependent difference maps to the observed difference maps[4]. Time-resolved absorption spectra were collected on crushed crystals from 20 µs to a few seconds at 0 °C and 30 °C with a home-built fast microspectrophotometer and analyzed also by SVD using MatLab (MathWorks) routines.

## Results:

Fig. 2A shows an example of right singular vectors (rSVs) extracted from the TRX data by SVD[11]. A global fit with four exponentials identifies four kinetic processes with relaxation times $\tau_1 \ldots \tau_4$ . The process with relaxation time $\tau_1$ results from the non-zero laser pulse width and the decay of $I_T$ to both $I_{CT}$ and $pR_1$ that can be identified in the earliest difference maps. Processes $\tau_2$ to $\tau_4$ result from relaxations of states $I_{CT}$ to $pR_2$ ($\tau_2$), the joint relaxations of $pR_1$ and $pR_2$ to pB and pG ($\tau_3$) and finally from pB to pG ($\tau_4$), respectively. These processes accelerate when the temperature is increased (see Fig. 2A, red numbers, and Fig. S5). In Fig. 2 B-D, the relaxation rates $\Lambda_i = 1/\tau_i$, are plotted as a function of temperature and fitted by the Van't Hoff-Arrhenius equation (dashed lines). The pre-factors and the energies of activation $E_a$ derived from the fits are shown in Tab. 1.

Slower relaxation times from TRX and time-resolved micro-spectroscopy (TRS) agree reasonably (Fig 3 and Tab. 1). At 0 °C three processes are observed. Process (1) corresponds to the pR to pB transition. The relaxation time derived from TRS is 2.1 ms which compares to process $\tau_3$ (0.7 ms) obtained from TRX. The pB to pG transition is biphasic (processes 2 and 3). The relaxation time of process (2) is 67 ms at 0 °C compared to $\tau_4$ = 73 ms obtained from TRX, and 2.2 ms at 30 °C compared to $\tau_4$ = 6 ms observed crystallographically. Process (3) observed with TRS at 0 °C contributes in a minor way, so that it cannot be detected with TRX. However, at elevated temperatures it can also be detected by crystallography with a similar relaxation time. Notably, the photocycle can be observed crystallographically up to 70 °C, and PYP also remains active at low temperatures (-40 °C) where the photocycle completes in ~15 s.

The main reaction pathway through $pR_2$ follows microscopic rate coefficients $k_1$, $k_3$, $k_4$ and $k_8$ (Fig. 4). A minor pathway involves $pR_1$ branching from $I_T$. Rate coefficient $k_7$ is generally 50% of $k_5$. One third of the $pR_1$ molecules relax directly to pG, two thirds populate pB. Since the $pR_1$ occupancy is low, the rate coefficient $k_7$ is difficult to determine and therefore may vary substantially. $pR_2$ typically decays mainly to pB, $k_6$ is generally much smaller than $k_4$. In earlier PYP studies[15,22] the pathways through $k_6$ and $k_7$ were not taken into account. Here we considered them as well, as more general possibility. Both pathways add little to the mechanism, since the $pR_1$ occupancy is small, and only a small fraction of $pR_2$ relaxes to pG. The main product is pB, which decays with $k_8$. At higher temperatures only rate coefficients $k_4$ to $k_{10}$ were included. This mechanism lacks the early intermediates. It features in addition two scale factors that account for the amount of $pR_1$ and $pR_2$ and an extra state $pB_2$. The two pR states are fully occupied after 20 ns at these temperatures[12] and relax to pB on the µs time scale. At +50 °C a weak second pB phase appears that indicates the presence of $pB_2$ (see Fig. S5). Rather than speeding up, the photocycle slows down (Fig. 2G) because the PYP occupies additional pB-like states even in the crystal. The concentration profiles of the intermediates closely display the relaxation times observed in the rSVs (Fig. S5).

**Table 1.** Energetics of the PYP photocycle. Upper: macroscopic, observable rates $\Lambda_i$ from TR crystallography. Temperature dependences are fit by the Van't Hoff-Arrhenius equation. The temperature dependence of process $\tau_1$ ($\Lambda_1$) cannot be determined due to limited time resolution. Middle: energetics derived from fitting the TSE to the temperature dependence of selected microscopic rate coefficients (errors from the fit in brackets). Lower: comparison of processes (1) to (3) observed by TRS with those derived from TRX ($\tau$3 and $\tau$4). An extra phase (3) is observed by TRX only at elevated temperatures.

| Macroscopic rate coefficients | $\Lambda_1$ | $\Lambda_2$ | $\Lambda_3$ | $\Lambda_4$ |
|---|---|---|---|---|
| Pre-factor $\nu$ [1/s] | n.a. | $1.6 \times 10^{14}$ | $1.9 \times 10^{13}$ | $7.7 \times 10^{10}$ |
| Energy of activation $E_a$ [kJ/mol] | n.a. | 35.9 | 53.4 | 49.6 |
| **Microscopic rate coefficients** | $^a k_3$ | $^a k_4$ | $^a k_5$ | $k_8$ |
| $\Delta H^\#$ [kJ mol$^{-1}$] | 24.7 | 48.1 | 50.0 | 75.2 (0.03) |
| $\Delta S^\#$ [J mol$^{-1}$ K$^{-1}$] | -9.4 | -10.2 | -14.8 | 41.9 (0.08) |
| $^b T\Delta S^\#$ [kJ mol$^{-1}$] | -2.8 | -3.1 | -4.5 | 12.6 (0.02) |
| $^b \Delta G^\#$ [kJ mol$^{-1}$] | 27.5 | 51.2 | 54.5 | 62.7 (0.09) |
| **Processes observed** | (1) \| $\tau_3$ | (2) \| $\tau_4$ | (3) | |
| 0 °C TRS | 2.1 ms | 67 ms | 800 ms | |
| TRX | 0.7 ms | 70 ms | n.o. | |
| 30 °C TRS | n.a. | 2.2 ms | 32 ms | |
| TRX | 99 µs | 6 ms | 40 ms (50 °C) | |

[a] errors of fit parameters smaller than 1%
[b] at 300 K
n.a. = not applicable, n.o. = not observed

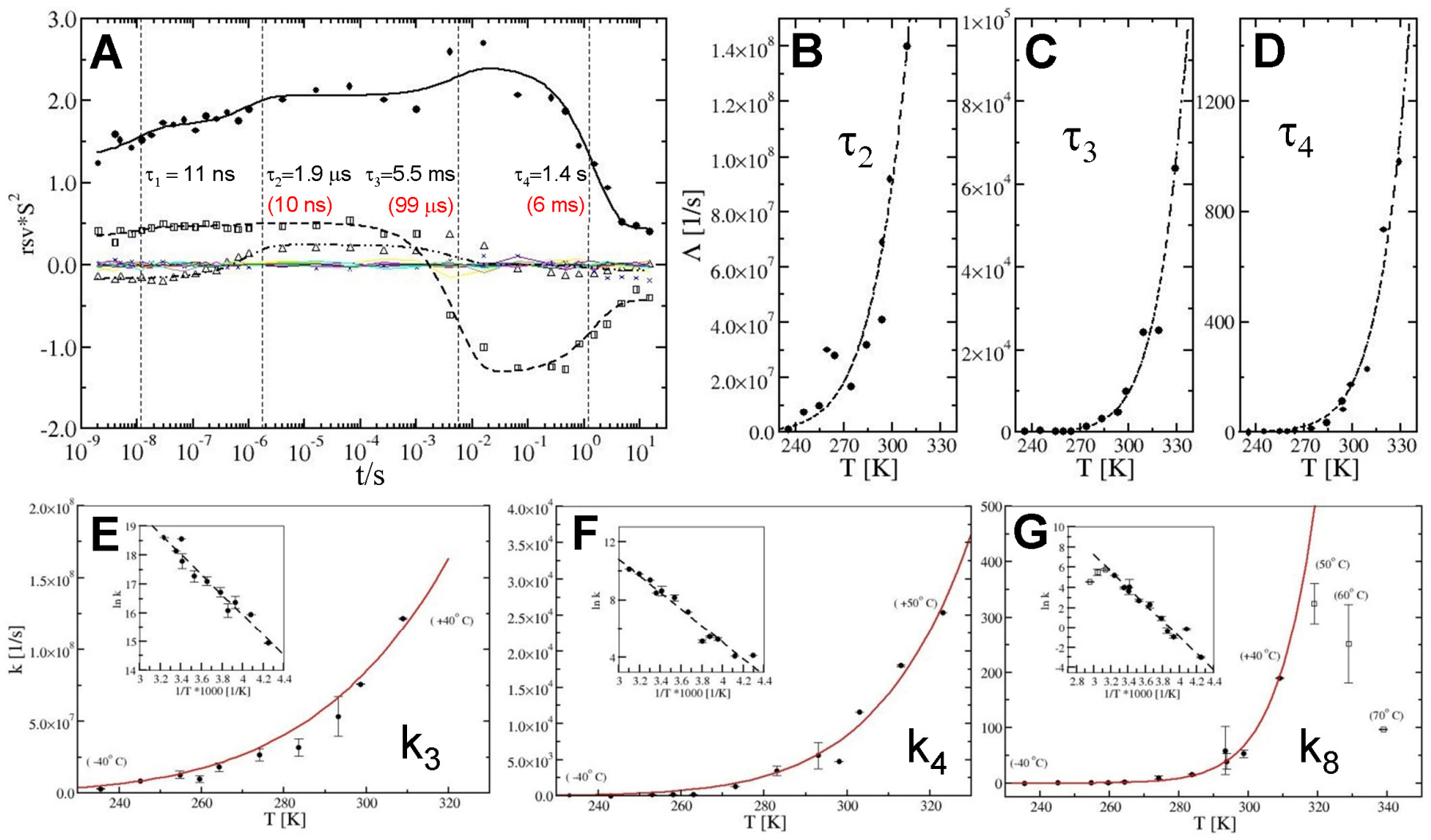

**Figure 2.** A: Right singular vectors (rSV) at -40 °C. The rSV are weighted by the square of their respective singular value S. Four kinetic processes $\tau_1$ …$\tau_4$ are globally observed (dashed vertical lines). Solid spheres, open triangles, squares: first, second and third significant rSVs. Colored thin lines around zero: less significant rSVs. Solid black line, dashed line, dashed double dotted line: global fit of the significant rSVs by four exponential functions, respectively, with the same set of relaxation times but different amplitudes. Relaxation times obtained at room temperature (25 °C, red) are shown for comparison in brackets. B-D: macroscopic rates $\Lambda$ (inverse of relaxation times) for processes $\tau_2$ to $\tau_4$ plotted as function of temperature, respectively. Dashed lines: fits by the Van't Hoff-Arrhenius equation. E-G: temperature dependence of the main-pathway microscopic rate coefficients $k_3$, $k_4$ and $k_8$; red lines: fits by the TSE. Inserts: Arrhenius plots; dashed lines: fits by straight lines.

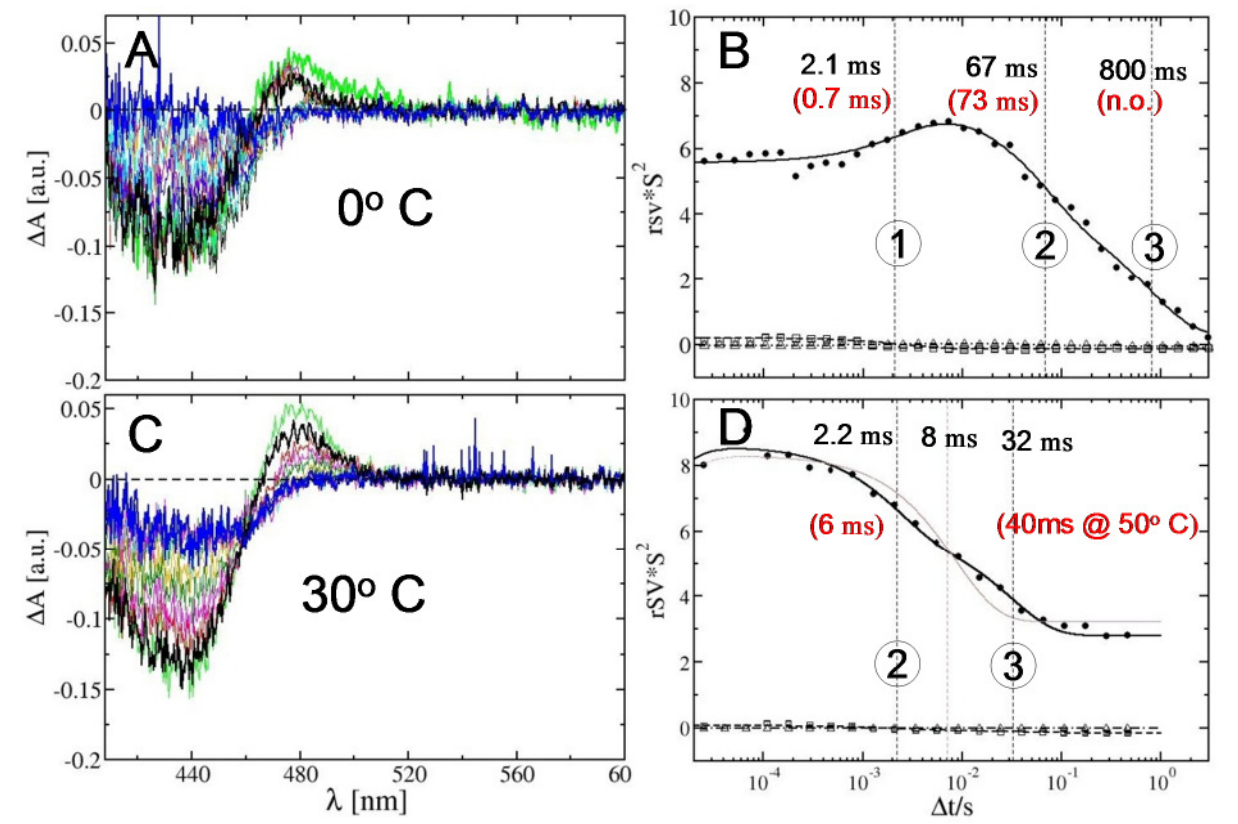

**Figure. 3** Time-resolved difference absorption spectra on crystalline slurry at various temperatures. A: 0 °C and C: 30 °C. (black: earliest time; blue last time-point). B and D right singular vectors (rSVs), respectively, from SVD analysis of the difference spectra. Solid lines: global fit by sums of three exponentials identifying 3 processes. Processes 1-3 are labeled with their respective relaxation time, and marked with thin vertical lines; relaxation times from crystallography are shown in red brackets (n.o.: not observed). Gray line in panel D: fit of the final relaxation with only one exponential.

The kinetic mechanism in Fig. 4 was used to extract time and temperature dependent rate coefficients, $k_i$, from the TRX data[12,21]. For all rate-coefficients, except $k_1$ and $k_2$ where we are limited by the time-resolution, thermal activation is observed. In Fig. 2E-G the temperature dependences of $k_3$, $k_4$ and $k_8$ are shown (those of $k_5$ and $k_7$.are shown in the SI). The barrier height varies from 27.5 kJ/mol for the $I_{CT}$ decay ($k_3$) to 63 kJ/mol for the $pB_1$ depopulation ($k_8$) (Tab. 1). Absorption of the laser pulse deposits energy into a volume approximately determined by the laser footprint on the crystal, the crystal diameter and the penetration depth. An adiabatic temperature jump of about 11 °C is estimated (SI). By correcting for the jump, the barriers shift only by ~1.5 kJ/mol, and the entropic contributions remain almost the same. A portion of the absorbed energy is initially stored in the twisted chromophore geometry of $I_T$, and is only released gradually through the exothermic relaxation processes. Due to the shallow penetration depth of the laser, the heat dissipates rapidly on a sub-ms time scale. The final pB to pG relaxation is never affected. Accordingly, we report here uncorrected values.

## Discussion:

The TSE describes the temperature dependence of the $k_i$ in the range from -40 °C up to 50 °C. This makes it possible to infer the free energy landscape of the PYP photocycle (Fig. 1). We adopted some of the conformational free energies from solution[23,24] with the strong caveat that they might be different in the crystal. Barriers of activation from TRX are shown for the main reaction pathway. The final barrier is the rate limiting step of the reaction. It slows down the photocycle decisively so that the signaling state may persist. PYP's chromophore pocket opens to the solvent and becomes exceptionally susceptible for additional stimuli such as the pH[22,25].

The TSE allows for the separation of entropic and enthalpic contributions to the barriers from the temperature dependent $k_i$ (Fig. 1). PYP intermediates occupy minima on the free energy surface, because the chromophore is immersed into a tight hydrogen bonding network. $\Delta H^{\#}$ is positive, because for the reaction to occur, some of the bonds have to be broken. $\Delta S^{\#}$ reflects the gain or loss of degrees of freedom. If the protein has time to relax, the structure can fluctuate through the substates and the entropy change is positive[26]. If, however, the protein environment stays rigid, there might be only one well defined, narrow path for the transition. In this case $\Delta S^{\#}$ may become negative. By inspecting the structures of the intermediates the structural reasons for the observed values become clear, although due to principal reasons, the structures of the transition states themselves remain unknown. For the $pR_2$ to pB transition, for example, the chromophore lifts out of a hydrogen bonding network involving Tyr42, Glu46 and Cys69. $\Delta H^{\#}$ is positive. Then, the chromophore rotates. On this time scale protein relaxations are incomplete. The difference maps are clean except in the direct vicinity of the chromophore (Fig. S2 B). Consequently, there is only limited space for this rotation and the entropic contribution to the barrier is slightly negative. The situation is different for the pB to pG transition. Once the hydrogen bonds of the pCA head hydroxyl to Arg52 and to one or two water molecules break, the pCA can re-isomerizes back to *trans*. At these elongated times the protein structure is relaxed as obvious from numerous difference electron density features on protein moieties surrounding the chromophore pocket (Fig. S2C). Once the pCA head is free, it can occupy an enlarged chromophore pocket. $\Delta S^{\#}$ becomes positive (Tab. 1). The entropic contribution helps to accelerate the reaction. In solution the PYP structure relaxes even further. For the pB to pG transition $\Delta H^{\#}$ is only 10 kJ/mol at pH 3[27] and $T\Delta S^{\#}$ is largely negative at -60 kJ/mol[27]. In solution the reaction is slowed down almost entirely due to the entropy and PYP refolds via an ordered moiety, whose structure remains elusive. In the crystal, however, the entropy plays a smaller role, because a highly disordered intermediate does not form and the transition state is more disordered than the pB state.

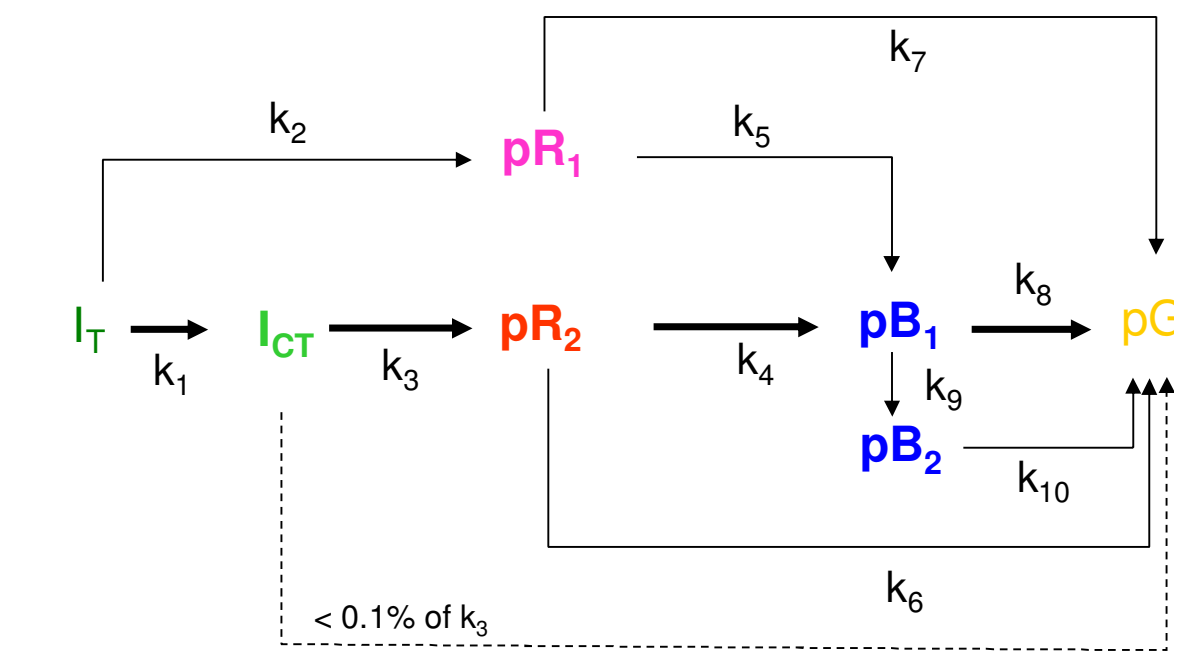

**Figure 4**. Chemical kinetic mechanisms and rate coefficients of the PYP photocycle. At temperatures up to 40 °C, 8 rate coefficients ($k_1$ to $k_8$) and 5 intermediate states, plus the dark state contribute. The occupation of $pB_2$ is very low and cannot be observed. Above 40 °C the early intermediates $I_T$ and $I_{CT}$ are not detected because the time-series start around 100 ns. $pB_2$ accumulates to detectable extent and rate coefficients $k_9$ and $k_{10}$ contribute in addition. The main reaction pathway in PYP is indicated by bold arrows. The direct path from $I_{CT}$ to pG is irrelevant (dashed arrow).

The PYP photocycle is an excellent model to study macromolecular reactions. Barriers of activation in the photocycle can now be determined from time-resolved crystallographic data alone. Enzymatic reactions are more difficult to investigate because they are non-cyclic or irreversible. However, free energy landscapes including barriers of activation are decisive for their function. It is therefore desirable to develop methods to routinely investigate these reactions by TR methods[28]. Our experiments guide the way to explore the vast phase-space available to proteins and enzymes using pulsed X-ray sources such as the synchrotron and emerging sources including X-ray free electron lasers[28-30].

**Supporting Information**. Detailed methods, difference electron density maps, spectra, and results from the SVD and kinetic analyses at all 14 temperature settings (Fig. S1 – Fig. S5) are presented as Supporting Information (SI). This material is available free of charge via the Internet at http://pubs.acs.org.

## AUTHOR INFORMATION

### Corresponding Author

Marius Schmidt, email: m-schmidt@uwm.edu, phone: 414-229-4338.

### Funding Sources

MS is supported by NSF CAREER grant 0952643. HI is supported by the Research Center Program (CA1201) of IBS (Institute for Basic Science) in Korea. Use of BioCARS was supported by the National Institutes of Health, National Institute of General Medical Sciences grant P41GM103543 (formerly National Center for Research Resources P41RR007707). The time-resolved set-up at Sector 14 was funded in part through a collaboration with Philip Anfinrud (NIH/NIDDK). Use of the Advanced Photon Source, an Office of Science User Facility operated for the U.S. Department of Energy (DOE) Office of Science by Argonne National Laboratory, was supported by the U.S. DOE under Contract No. DE-AC02-06CH11357.

## ABBREVIATIONS

pCA, para-coumaric acid; PYP, photoactive yellow protein; SVD, singular value decomposition; TRS, time-resolved spectroscopy; TRX, time-resolved X-ray crystallography.

# Protein Energy Landscapes Determined by 5-Dimensional Crystallography

Marius Schmidt[*,†], Vukica Srajer[‡], Robert Henning[‡], Hyotcherl Ihee[∥], Namrta Purwar[†], Jason Tenboer[†], Shailesh Tripathi[†]

[†]Physics Department, University of Wisconsin-Milwaukee, Milwaukee, WI, USA,

[‡]Center for Advanced Radiation Sources, The University of Chicago, Chicago, IL, USA

[∥]Nanomaterials and Chemical Reactions and Institute for Basic Science, Daejeon, 305-701, Republic of Korea and Department of Chemistry, KAIST, Daejeon, 305-701, Republic of Korea

## Supporting Information

### Methods

***Crystals and Data Collection*:**

PYP crystals were grown as described elsewhere[31]. Crystals were mounted in 1.0 mm glass capillaries at pH 7. The glass capillaries were glued to brass pins to be mounted on the goniostat. For experiments above 30°C, the capillaries must be insulated from the brass pins as shown in Fig. S1, Otherwise severe distillation effects due to the cold brass pins will dry out the crystals. Typical crystal sizes used were 100 x 100 x 700 μm$^3$. The temperature was controlled from -40 °C to 20 °C by a Cryojet (Oxford Instruments/Agilent Technologies) and from 25 °C to 70 °C by a Cryostream (Oxford Cryosystems) aligned coaxially to the capillary. The temperature at the crystal position was determined by a calibrated thermocouple. To initiate the photocycle a ns laser Opolette HEII (Opotek) was used. Laser pulses of 7ns duration (FWHM) at 485nm were transported by an optical fiber. The laser light was focused at the crystal to a round focal spot of 200 μm with typical pulse energy densities from 4.0 to 4.5 mJ/mm$^2$. The reaction was followed by a series of Laue diffraction snapshots at various time delays between the ns pump (laser) and 100ps probe (X-ray) pulses. Up to 40 °C, time delays ranged from 2 ns to several seconds on an equidistant logarithmic time scale. From 50 °C and higher, delay times started at 100 ns (Tab. S1). Depending on the crystal size, three to seven pump-probe pulse sequences were accumulated prior to detector readout to obtain high-quality diffraction patterns. The waiting time between the pulse sequences, necessary for the dark state recovery, varied between 1 s at higher temperatures and around 20 s at lowest temperatures (Tab. S1). Single 100ps X-ray pulses were isolated as described[32]. The X-ray beam was focused to a size of 60 μm vertically (v) and 90 μm horizontally (h) and each 100ps pulse contained about $4 \times 10^{10}$ photons in the hybrid mode of the APS storage ring or $10^{10}$ photons in the 24-bunch mode. At a particular crystal orientation we only probed by X-rays the crystal surface layer that was facing the laser[32]. To precisely position this layer in the X-ray beam, crystal edge scan was done where a series of weak diffraction images were collected while the crystal edge was translated through the X-ray beam. With this, the overlap of the laser-excited volume with the X-ray probed volume is optimized. Diffraction images were recorded for the complete series of time delays using the same volume of the crystal; the crystal was then rotated and translated to a new position to collect the diffraction images for the same time series. This way reciprocal space was covered by 20 different crystal orientations with time as the fast variable and crystal orientation as the slow variable[11]. By translating the crystal along its axis for each of these 20 different orientations, a fresh volume of the crystal is exposed to minimize radiation damage[33].

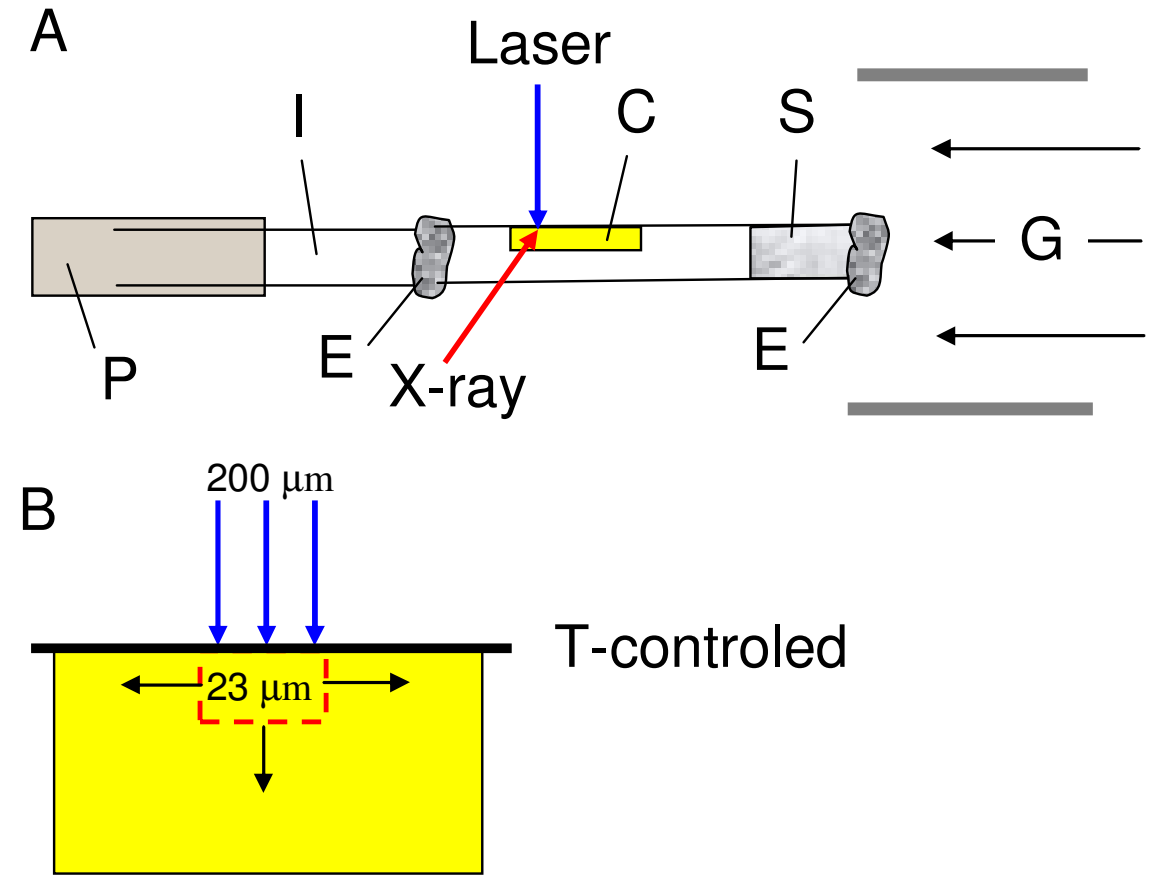

**Figure S1.** A: Capillary mount at T > 30° C. P: brass pin, I: insulating glass capillary, E: epoxy glue, C: PYP crystal, S: stabilizing solution. G: temperature controlled gas stream. B: laser illuminated volume of the crystal (red dashed box). Capillary wall (thick black line) is temperature controlled. Black arrows: heat diffuses out.

**Table S1.** X-ray data

| T [K] | time-points | from to | wait time between laser pulses |
|---|---|---|---|
| 235.5 | 31 | 2 ns to 15 s | 20 s |
| 245.1 | 27 | 1 ns to 4 s | 12 s |
| 254.7 | 34 | 2 ns to 8 s | 12 s |
| 259.5 | 29 | 2 ns to 8 s | 10 s |
| 264.3 | 23 | 4 ns to 8 s | 10 s |
| 274.0 | 27 | 8 ns to 512 ms | 10 s |
| 283.6 | 27 | 2 ns to 128 ms | 4 s |
| 293.2 | 27 | 2 ns to 128 ms | 4 s |
| 293.7 | 27 | 2 ns to 512 ms | 2 s |
| 298.5 | 27 | 2 ns to 128 ms | 2 s |
| 308.9 | 24 | 2 ns to 64 ms | 2 s |
| 318.9 | 22 | 100 ns to 256 ms | 4 s |
| 328.8 | 24 | 100 ns to 1 s | 4 s |
| 338.8 | 21 | 100 ns to 1 s | 4 s |

***Laue Data Processing***:

Laue data were processed with Precognition and Epinorm (RenzResearch). Weighted difference structure factor amplitudes were calculated as described[22,34]. From these and the phases of the dark state PYP model from the protein data bank (PDB)[35] entry 2PHY, time-and now temperature-dependent difference electron density maps ($\Delta\rho^{obs}$) were calculated. Hence, the experiment is represented by a 5-dimensional array of crystallographic data.

***Macroscopic Relaxation Times***:

The time-series of the difference maps were analyzed at each temperature by singular value decomposition (SVD) as described[4,11]. The spatial components, which are difference electron densities, are separated into the left singular vectors (lSVs). The kinetics can be found in the significant right singular vectors (rSVs). Sums of exponential functions with different amplitudes but with the same set of characteristic times were globally fit to the set of significant rSVs. These characteristic times, τ, are the observable, macroscopic relaxation times of the kinetics. The macroscopic rates Λ (inverse of relaxation times) were plotted as a function of temperature and fitted by the Van't Hoff-Arrhenius equation (see text).

***Chemical Kinetic Mechanism and Microscopic Rate Coefficients***:

Kinetic modeling is required to explain the time-dependent variations of the difference electron density values in terms of concentrations of the intermediates. All structures of the PYP photocycle intermediates in the time-range analyzed are known[12,15,21,31]. Each structure represents a transient state. Up to $40^o$ C the states are arranged in the chemical kinetic mechanism shown in Fig. 4 based on known intermediates and including all forward-going transitions. At $50^o$ and higher the fast processes are not reliably observed. The mechanism lacks the early intermediates $I_T$ and $I_{CT}$, and it features in addition state $pB_2$ that appears as an additional phase in the rSVs. The structures are, in the order of appearance after reaction initiation: $I_T$, $pR_1$, $I_{CT}$, $pR_2$, $pB_1$ and $pB_2$[12,15,21]. Finally, the dark (reference) state pG is recovered. The atomic coordinates of states $I_T$ and $I_{CT}$ were obtained from a recent publication[12] with protein data bank-entries 3VE3 and 3VE4, respectively. Those of all other states were retrieved from the protein data bank entries 2PHY (dark/reference), 1TS7 ($pR_1$ and $pR_2$), 1TS0 ($pB_1$) and 1TS6 ($pB_2$). The microscopic rate coefficients $k_i$ connect the states (Fig. 1). The time-dependent fractional concentrations $c_i(t,k)$ of the intermediates are linear combinations of the rate coefficients and are calculated by solving the coupled differential equations describing the mechanism in Fig. 4[4,36]. The concentrations are then used to calculate time-dependent difference electron density maps from the time-independent difference maps of the intermediates. For this, first, structure factors derived from the dark state model pG are subtracted from those of the models representing the intermediate states. From these, the time-independent difference maps of the intermediates $\Delta\rho_i^{ind}$ are calculated. Using the concentrations, these maps are finally linearly combined to the time-dependent difference maps ($\Delta\rho^{calc}$), which can be compared to the observed, time dependent difference maps $\Delta\rho^{obs}$ at each particular temperature:

$$\Delta\rho^{calc} = \sum_{i=1}^{N} c_i(t,k)\Delta\rho_i^{ind} \tag{S1}$$

The electron density features typically follow exponential functions with amplitude $A_i$ and characteristic times $\tau_i$ (relaxation times, or their reciprocals, relaxation rates $\Lambda_i$), each for a particular intermediate i. Both, the amplitudes and the relaxation times are linear combinations of the microscopic rate coefficients of the underlying chemical, kinetic mechanism[1,36]. These relaxation times are also observed globally as kinetic processes in the rSVs. However, if relaxation times are similar, they appear only as one process. This is why only 4 processes are observed in our rSVs, but 5 transient intermediates contribute. The amplitudes are equivalent to (fractional) concentrations, and the relaxation times determine the variations of the concentrations with time. On the absolute scale, difference electron densities are directly related to concentrations (Eq. S1). In certain instances[37] one can account for the electrons in a particular density feature and infer from this the occupation, the fractional concentration, or even the concentration proper, of a molecule in the unit cell. In other instances the entire difference electron density of the whole unit cell can be fit by calculated difference electron density maps from chemically plausible structural models. This is in contrast to absorption spectroscopy where there are linear factors between concentration and absorption, namely the absorption coefficients that are all a-priori unknown for the intermediate states[17,20,38]. Initially determined absorption coefficients will yield self-consistent results, and may persist in the literature. This danger is considerably smaller in crystallography. A chemical kinetic mechanism compatible with the X-ray data must produce time-dependent concentrations (amplitudes as well as well as relaxation times) that are directly commensurable with the observed difference electron density values. The mechanism becomes testable and, in certain instances falsifiable by posterior analysis [4,11,15]. Still, a number of mechanisms can fit the data reasonably well, and consequently are indistinguishable or degenerate[5].

Posterior analysis allows the refinement of the microscopic rate coefficients of a mechanism. When the microscopic rates k are varied, the concentrations of the intermediates $c_i(t)$ are changed. This in turn modifies the $\Delta\rho^{calc}$. Variation of the rates k is performed until optimal agreement with the measured $\Delta\rho^{obs}$ is achieved[21,22]: $\sum_t \left[\Delta\rho^{obs} - sf\cdot\Delta\rho^{calc}(k)\right]^2 \rightarrow \min$, where the differences from all t available time points in a time series are accounted for and sf is a (typically single) scale factor. This approach has the potential to exclude reaction pathways. This is the case when the refinement yields rate coefficients that are so small that they may be ignored. An example is shown in Fig. 4 (see main text), where the rate coefficient of the dashed pathway is less than 0.1% of $k_3$ throughout and, therefore, less than 0.1% of the molecules react through this pathway. The extent of reaction initiation is represented by a scale factor (sf) that is also varied freely in the fit. For the mechanism in Fig. 4, a single scale factor determines the initial concentration of state IT, for the mechanism at T ≥ 50 °C, two scale factors were used accounting for the initial concentrations of $pR_1$ and $pR_2$. In this way, at each temperature, a set of rate coefficients was determined. Up to -10° C rate coefficients $k_1$ and $k_2$ were free fit parameter. From 0 °C $k_1$ was fixed to $2 \times 10^9$ 1/s, and changed to $3 \times 10^9$ 1/s from 25 °C (Tab S2). $k_2$ was varied freely throughout, because at a given $k_1$, the magnitude of $k_2$ accounts for the concentration of $pR_1$ relative to $I_{CT}$. Since the concentration of $pR_1$ is directly observable in the difference maps, $k_2$ is determined. The temperature dependences of the individual rate coefficients were then determined by directly fitting the transient state equation (Eq. 1).

**Table S2.** Rate coefficients in [1/s] at the temperature range covered (na: not available).

| T jet | T diode | $k_1$ | $k_2$ | $k_3$ | $k_4$ | $k_5$ | $k_6$ | $k_7$ | $k_8$ | $k_9$ | $k_{10}$ |
|---|---|---|---|---|---|---|---|---|---|---|---|
| -40 °C | 235.5 K | $2.3 \times 10^7$ | $8.3 \times 10^6$ | $3.07 \times 10^6$ | 63.9 | 35.9 | 22.8 | 3.3 | 0.05 | 0 | 0 |
| -30 °C | 245.1 K | $9.9 \times 10^7$ | $7.5 \times 10^7$ | $8.45 \times 10^6$ | 61.0 | 7.81 | 25.1 | 2.7 | 0.84 | 0 | 0 |
| -20 °C | 254.7 K | $3.8 \times 10^8$ | $1.3 \times 10^8$ | $1.29 \times 10^7$ | $1.94 \times 10^2$ | 7.09 | 38.2 | 7.3 | 0.36 | 0 | 0 |
| -15 °C | 259.5 K | $4.5 \times 10^8$ | $3.3 \times 10^7$ | $9.84 \times 10^6$ | $2.30 \times 10^2$ | 81.4 | 32.2 | 26.7 | 0.77 | 0 | 0 |
| -10 °C | 264.3 K | $6.0 \times 10^8$ | $3.0 \times 10^8$ | $1.82 \times 10^7$ | $1.68 \times 10^2$ | 120 | 29.9 | 51.9 | 2.56 | 0 | 0 |
| 0 °C | 274.0 K | $2.0 \times 10^9$ | $6.0 \times 10^8$ | $2.67 \times 10^7$ | $1.33 \times 10^3$ | 122 | 159 | 173 | 8.07 | 0 | 0 |
| 10 °C | 283.6 K | $2.0 \times 10^9$ | $9.9 \times 10^8$ | $3.18 \times 10^7$ | $3.44 \times 10^3$ | 802 | 287 | 515 | 14.4 | 0 | 0 |
| 20 °C | 293.2 K | $2.0 \times 10^9$ | $1.3 \times 10^9$ | $5.34 \times 10^7$ | $5.59 \times 10^3$ | 2226 | 198 | 296 | 62.9 | 0 | 0 |
| 25 °C | 293.7 K | $3.0 \times 10^9$ | $1.8 \times 10^9$ | $1.16 \times 10^8$ | $4.78 \times 10^3$ | 4780 | 79.0 | 1100 | 39.1 | 0 | 0 |
| 30 °C | 298.5 K | $3.0 \times 10^9$ | $1.3 \times 10^9$ | $7.59 \times 10^7$ | $1.16 \times 10^4$ | 3440 | 605 | 4340 | 53.3 | 0 | 0 |
| 40 °C | 308.9 K | $3.0 \times 10^9$ | $1.5 \times 10^9$ | $1.22 \times 10^8$ | $1.80 \times 10^4$ | 3040 | 215 | 3850 | 190 | 0 | 0 |
| [a]50 °C | 318.9 K | - na - | - na - | - na - | $6.7 \times 10^4$ | $2.8 \times 10^4$ | 600 | $1.8 \times 10^4$ | 79.8 | 250 | 1.5 |
| [d]60 °C | 328.8 K | - na - | - na - | - na - | $6.58 \times 10^4$ | $4.28 \times 10^4$ | 600 | $1.0 \times 10^4$ | 326 | 138 | 24 |
| [d]70 °C | 338.8 K | - na - | - na - | - na - | $2.03 \times 10^5$ | $1.04 \times 10^5$ | - na - | - na - | 97.5 | 0 | 0 |

[a] At 50 °C and 60 °C $pB_1$ decays with $k_8$+$k_9$

[d] at T = 70 °C data quality is poor (see Fig. S5). Therefore, formation and decay of only one pB state is listed.

If N intermediates contribute to a time-course of crystallographic data, N amplitudes of difference electron densities and N relaxation times can be observed, hence there are 2N observables. At lower temperatures $pB_2$ is not observed and 8 microscopic rate-coefficients ($k_1$ ... $k_8$) are fit parameters in the mechanism that connects 5 intermediate states and the dark state. One additional fit parameter is the extent of reaction initiation. Hence, there are 9 free parameters. Since the time-course contains 5 intermediate states, the number of observables is 10. A least squares fit of the mechanism to the data is possible at all temperatures ≤ 40 °C. A ninth rate coefficients (a 10th fit parameter) spanning from $I_{CT}$ directly to pG can be included. However, this rate coefficient is smaller than 0.1% of $k_3$ so that the pathway is irrelevant within the constraints of the mechanism. Also $k_6$ is much smaller than $k_4$ throughout (see Tab. S2). As a consequence, the fitted values of $k_6$ are fluctuating largely and contribute little to the mechanism. This pathway can also be ignored. Above 40 °C a truncated mechanism is used that starts from the pR states, since the earlier processes become successively inaccessible at these temperatures. This mechanism includes in addition $pB_2$ because a second pB phase is identified in the rSVs. 8 observables are faced by 9 free parameters, which are the two scale factors that determine the extent of $pR_1$ and $pR_2$ and 7 rate-coefficients. In its full extent, this mechanism is underdetermined and cannot be fit without using additional constraints. Accordingly, we constrained the initial concentration of $pR_1$ to 30% of $pR_1$. In addition we fixed $k_6$ to 1% of $k_4$. Both conditions are roughly observed at lower temperatures. With this the fit becomes stable.

***Micro-spectrophotometry*:**

Small PYP crystals were crushed between two cover slides which were subsequently sealed with epoxy. The crystalline slurry was probed by a micro-spectrophotometer with a time-resolution of 20 μs. The design of the micro-spectrophotometer will be reported elsewhere. Briefly, the light of a 300W Xenon lamp (Ashai spectra) is collected into a tapered fiber. The fiber light is focused to about 200 μm onto the crystalline slurry. The transmitted light is picked up and transported to a Shamrock spectrophotometer equipped with a fast Andor Istar camera. To avoid bleaching by the intense monitoring Xenon light, a shutter (Thorlabs) restricts exposure to 13 ms. A reaction in the slurry is started by a 446 nm nanosecond laser pulse from the same Opolette II tunable laser which has been used also for the crystallographic experiments. The laser pulse energy density was about 50 μJ/mm² at the sample. The absorbance at 449 nm was around one. The laser flashlamp, the Q-switch, the shutters and the fast camera are synchronized by two Stanford Research Systems pulse generators. The experiment is controlled by a *MatLab* routine in conjunction with Andor's *Solis* kinetic software. The image intensifier of the Andor detector is time-gated at a time delay Δt after the laser pulse to collect a spectrum. Time-gate pulses are 20 μs for time-delays up to 100 ms and up to 700 μs for time-delays longer than 200 ms. This ensures that at any time delay the signal collected does not suffer from extensive dark noise of the CCD sensor. The photocycle was restarted approximately 25 times to collect an absorption spectrum with a sufficient signal to noise ratio. The temperature was controlled by an Oxford Cryojet HTII gas stream. Difference spectra were generated by subtracting the absorption spectrum collected in the dark from those collected at the time-delays. The time-series of difference spectra (Fig. 3, main text) were analyzed by singular value decomposition and relaxation times were extracted by fitting exponential functions to the right singular vectors in a similar way as it was done for the time-resolved crystallographic experiments.

***Estimation of energy and heat deposition as well as heat diffusion in the PYP crystal due to Laser pulses***:

The absorption coefficient of crystals is anisotropic[39]. With unpolarized laser light used in our experiments we can assume that the absorption coefficient of the absorption maximum is equal to that in solution (45,500 $cm^2$ $mmol^{-1}$ ), however slightly shifted (3 nm) in the crystal (Fig. S2). The wavelength of the laser used to initiate the reaction is 485 nm. At that wavelength the absorption coefficient is a factor of 10 smaller (see also Fig. S2). The PYP concentration is 96 mmol/L in the crystal. The penetration depth can be defined as the thickness when 1 a.u. is reached, hence 90% of the photons are absorbed. At 485 nm, 1 a.u. is reached at $d = \frac{1}{\varepsilon \cdot c} = 0.0023\,cm$ which is about half of the vertical size of 60 µm of the X-ray beam. The laser energy density in the focal spot at the crystal is around 4.0 $mJ/mm^2$. The focal spot has a diameter of 200 µm. The intersection of the laser beam with the crystal is 200 µm x crystal diameter which is about 120 µm. The fraction of the laser energy that strikes the crystal is therefore 0.2 mm × 0.12 mm × 4.0 $mJ/mm^2$ = 0.096 mJ. 90 % of this distributes to a volume of 0.2 × 0.12 × 0.023 $mm^3$ =5.5 × $10^{-10}$ L, which contains 5.3 × $10^{-8}$ mmol PYP. With a molecular weight of 14700 g/mol this amount of PYP has a mass of 8 x $10^{-10}$ kg. If we assume that half of the laser light is dissipated as heat into the vibrational modes of motion and half of it is stored in an energy rich chromophore configuration [18,24], we can estimate the adiabatic temperature rise from the fraction of the heat dissipated into the vibrational modes. With a heat capacitance of a typical protein of 5 kJ $kg^{-1}$ $K^{-1}$ [40], we would expect an increase of $0.5 \times 0.9 \times \frac{0.096 \times 10^{-6}\,kJ}{8 \times 10^{-10}\,kg \times 5kJ\,kg^{-1}K^{-1}} = 11K$ , which is about the typical temperature step of ~10 K we used. The energy stored in the chromophore configuration is gradually released. The total heat is diffusing out of the illuminated and excited volume. Since the penetration depth of the laser is much smaller than the crystal diameter, and the face of the crystal is in contact with the temperature controlled capillary surface we have a two dimensional heat diffusion problem with the heat escaping to the left and right into the crystal volume and into the capillary wall and into the bulk of the crystal volume below the illuminated volume. The characteristic time for the heat to diffuse out of that volume is [41]

$$\tau = \frac{1}{\kappa\pi^2\left(\frac{1}{a^2} + \frac{1}{b^2}\right)},$$

with κ the thermal diffusivity and a and b the size of the two-dimensional box shown in fig. S1. With a=0.2 mm, b=0.023 mm, and assuming that κ of the protein crystal equals to that of water being 0.143 $mm^2/s$, we estimate a characteristic time of roughly 0.5 ms for the heat to diffuse out of the laser illuminated volume. Moffat et al. [42] estimate heat diffusion times that are much longer based on calculations with larger illuminated volumes.

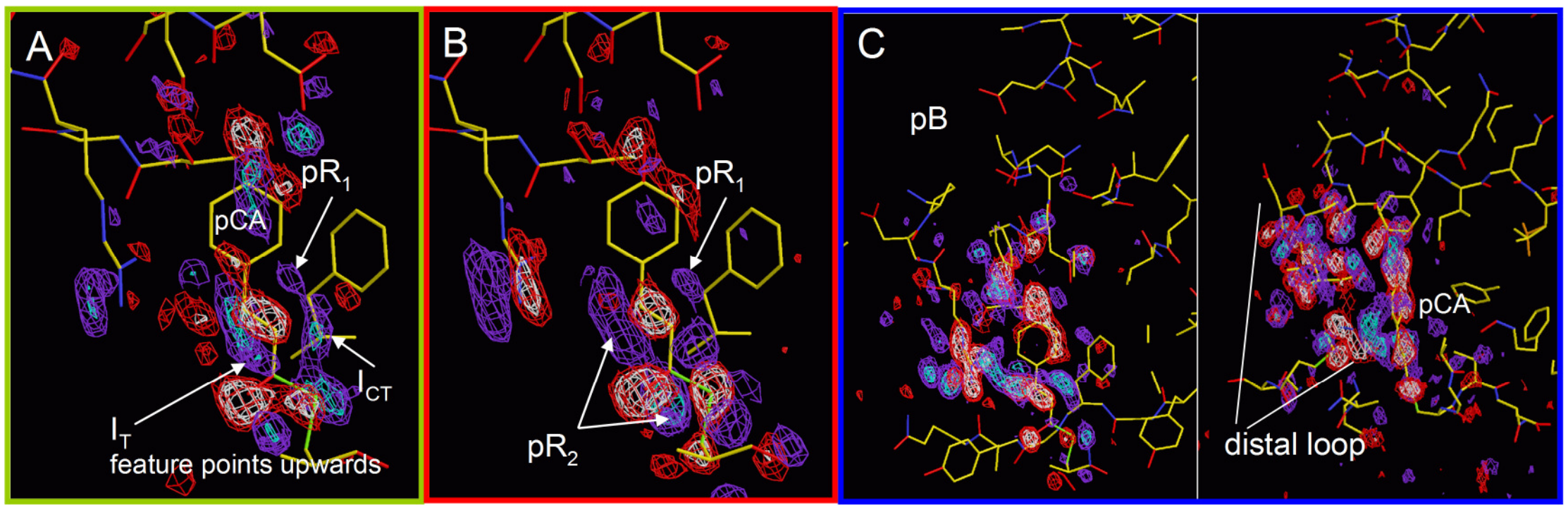

**Figure S2.** Difference maps near the p-coumaric acid (pCA) chromophore in PYP averaged through various time intervals and at various temperatures. A: average difference map obtained from the 8 earliest difference maps from 2 ns to 8 ns at T < -10° C, features of three intermediates $I_T$, $I_{CT}$ and $pR_1$ contribute to the same set of maps. B: average difference map obtained from 20 difference maps at T < -10° C from 256 ns to 64 µs, features of two intermediates, $pR_1$ and $pR_2$ contribute. C: average difference map obtained from 20 difference maps in the millisecond time range at 30° < T < 50° C. Main features are found near the chromophore and on the distal loop.

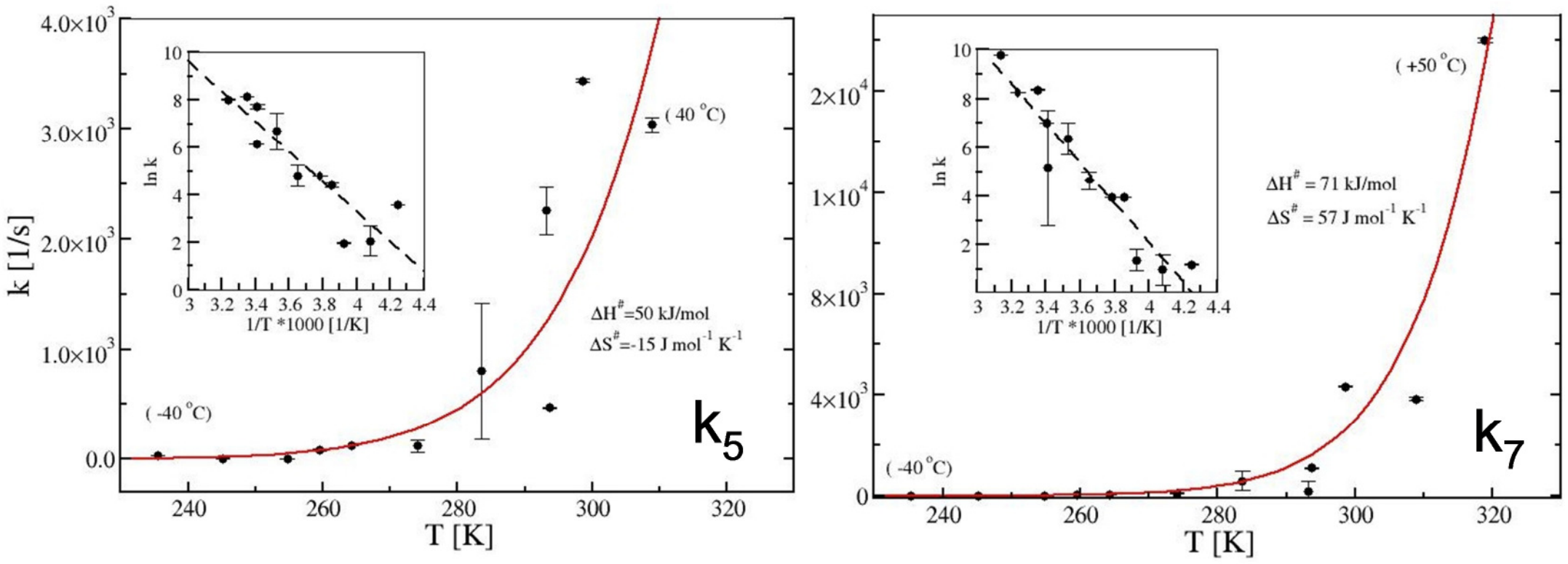

**Figure S3.** Temperature dependence of rate coefficients $k_5$ and $k_7$. Decay of $pR_1$ to $pB_1$ ($k_5$) and pG ($k_7$). Black spheres: rate coefficients, red line: fit by the transition state equation. Entropy and enthalpy of the barriers are listed. Inserts: corresponding Arrhenius plots.

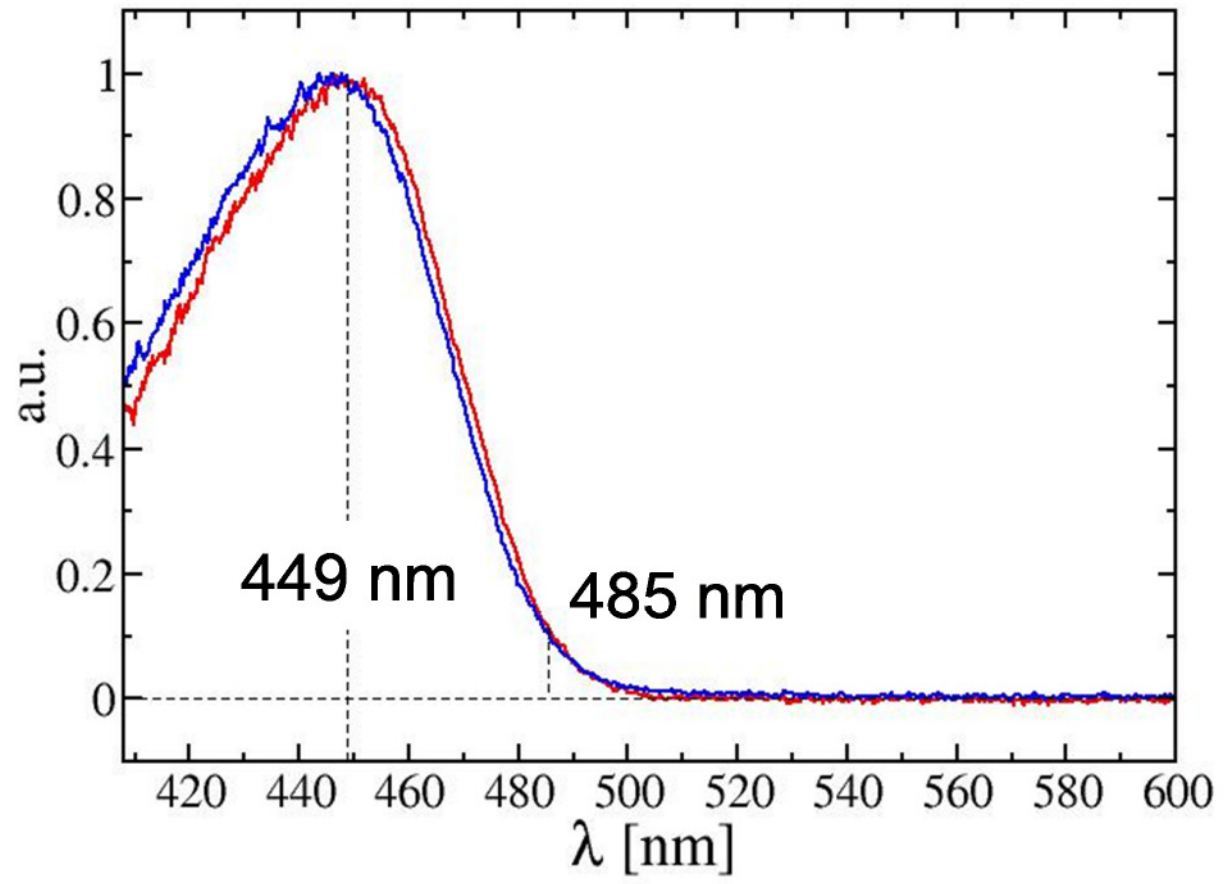

**Figure S4.** Typical PYP absorption spectrum as measured by the micro-spectrophotometer without laser illumination. Blue: solution, Red: crystalline slurry. Spectra normalized. Absorption maximum of crystal is shifted by ~ 3 nm as compared to solution. The absorption at 485 nm is a factor of ~10 smaller than at 449 nm.

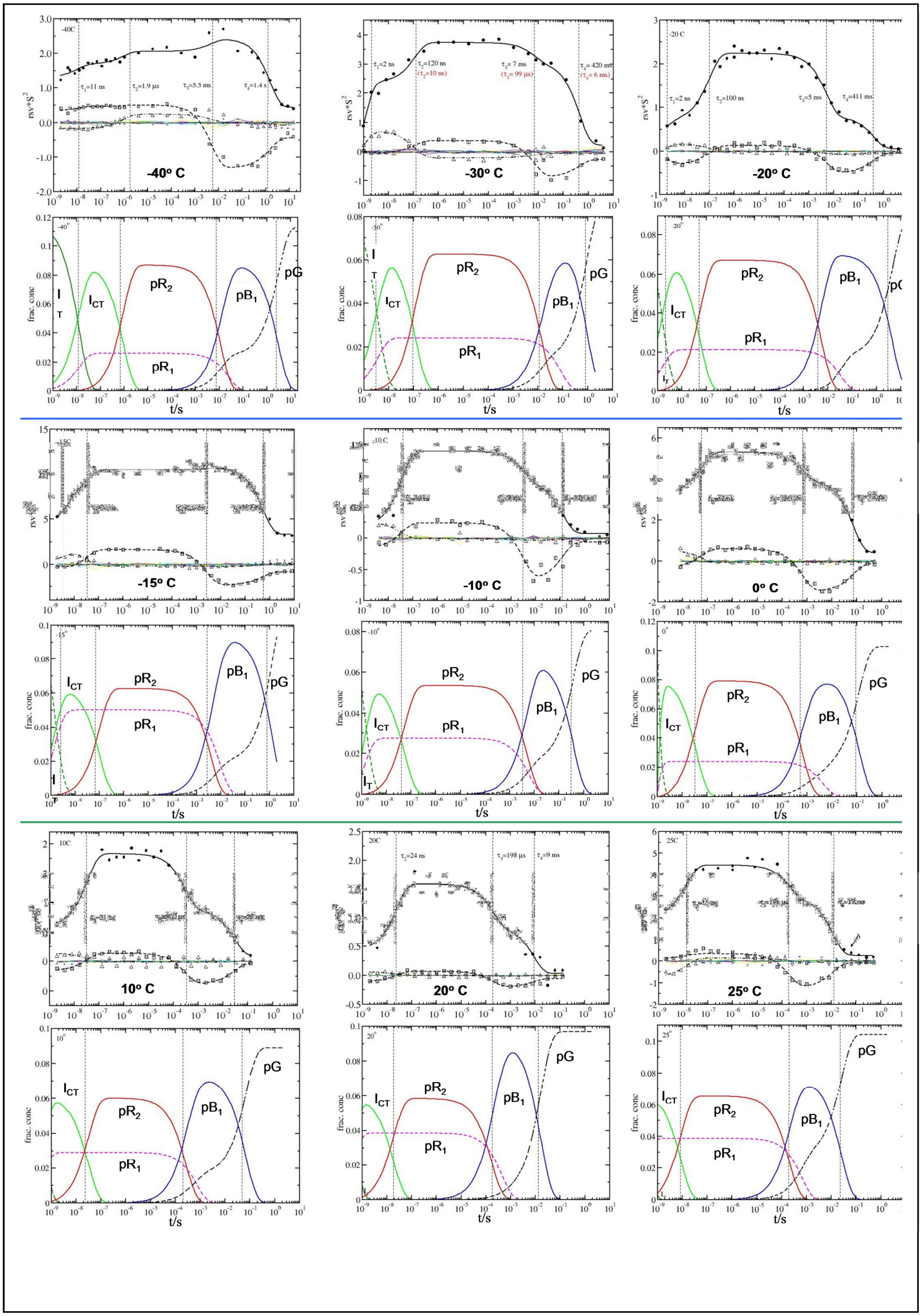

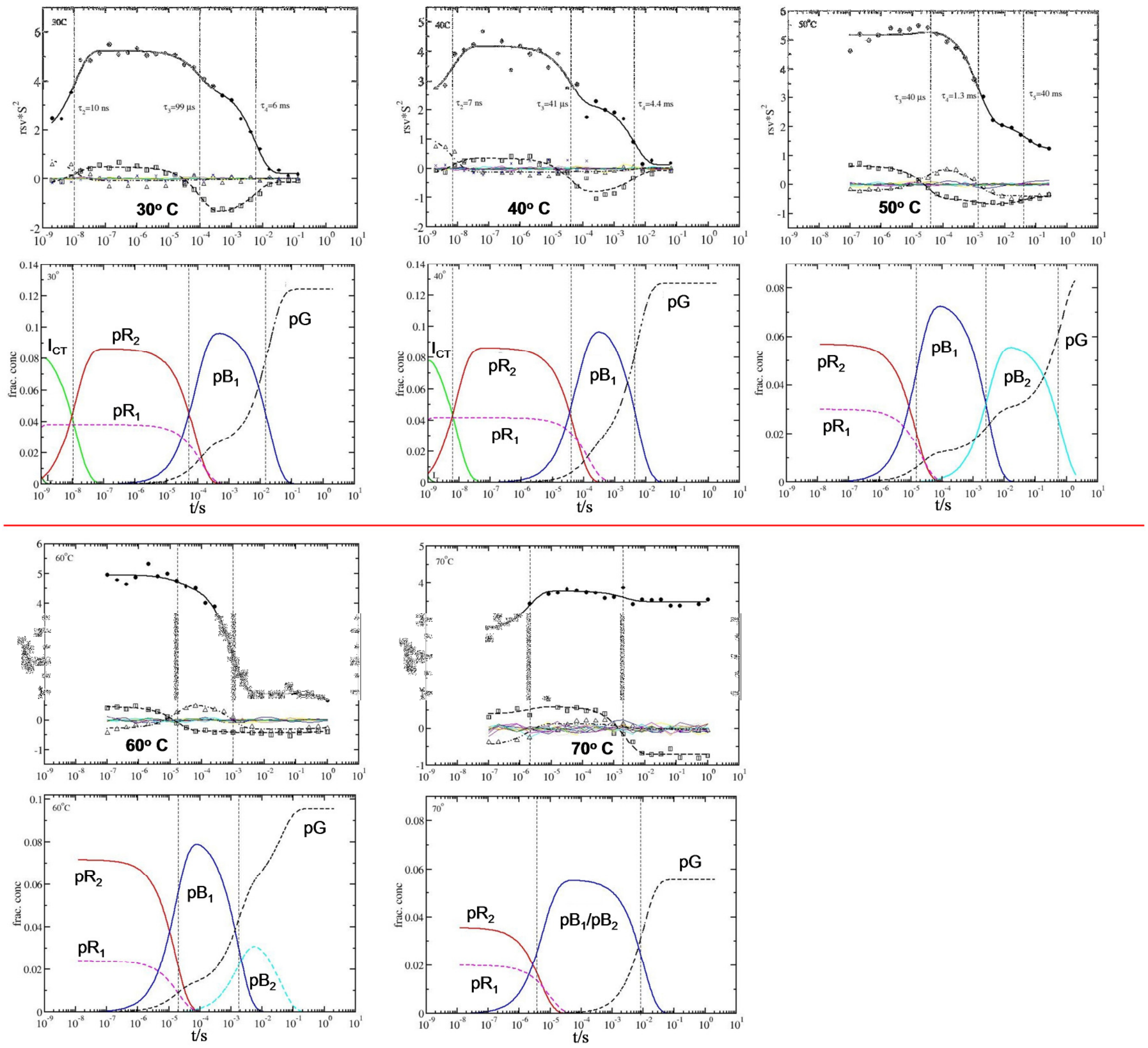

**Figure S5**. Right singular vectors (upper panels) as extracted at the 14 temperatures. Symbols denote the components of the right singular vectors multiplied with the square of the corresponding singular value. Black lines are from the global fit of a sum of exponentials. Solid spheres and black solid lines: first right singular vectors, open squares and black dashed lines: second right singular vectors, open triangles and black dashed dotted lines: third right singular vectors. Colored lines: less significant right singular vectors 1-15. Corresponding time-dependent fractional concentrations are shown in the panels below. Dark green: concentrations of $I_T$. The dashed dark-green color is used when $I_T$ decays within the pulse duration of the laser. Green: concentrations of $I_{TC}$. Red: concentrations of $pR_2$. Dashed magenta: concentrations of $pR_1$, this pathway is a minor pathway. Blue: concentrations of pB. Light blue: concentrations of $pB_2$. Dashed black: recovery of pG. Black vertical lines in upper and lower panels: comparison of relaxation times from the singular value decomposition and from posterior analysis. At T≥ 60 °C (below the thick red line), the offset in the rSV becomes excessively large, which indicates poor and noisy data. A kinetic analysis becomes challenging.